\documentstyle[12pt]{article}
\begin{document}

\def\beq{\begin{equation}}
\def\eeq{\end{equation}}
\def\bce{\begin{center}}
\def\ece{\end{center}}
\def\bea{\begin{eqnarray}}
\def\eea{\end{eqnarray}}
\def\ben{\begin{enumerate}}
\def\een{\end{enumerate}}
\def\ul{\underline}
\def\ni{\noindent}
\def\nn{\nonumber}
\def\bs{\bigskip}
\def\ms{\medskip}
\def\wt{\widetilde}
\def\tr{\mbox{Tr}\, }
\def\brr{\begin{array}}
\def\err{\end{array}}

\hfill UB-ECM-PF 94/34

\hfill November 1994

\vspace*{3mm}

\begin{center}

{\LARGE \bf
 Dilaton-Maxwell gravity with matter \hspace{2cm}  near two dimensions}

\vspace{8mm}

\renewcommand
\baselinestretch{0.5}
\medskip

{\sc E. Elizalde}
\footnote{E-mail: eli@zeta.ecm.ub.es} \\
Center for Advanced Study CEAB, CSIC, Cam\'{\i} de Santa
B\`arbara,
17300 Blanes
\\ and Department ECM and IFAE, Faculty of Physics,
University of Barcelona, \\ Diagonal 647, 08028 Barcelona,
Catalonia, Spain \\
 and \\
{\sc S.D. Odintsov} \footnote{E-mail: odintsov@ecm.ub.es.
On
leave from: Tomsk Pedagogical Institute, 634041 Tomsk, Russia.}
\\
Department ECM, Faculty of Physics,
University of  Barcelona, \\  Diagonal 647, 08028 Barcelona,
Catalonia,
Spain \\

\vspace{15mm}

{\bf Abstract}

\end{center}

Unlike Einstein gravity, dilaton-Maxwell gravity with matter is
renormalizable in $2+\epsilon$ dimensions and has a smooth $\epsilon
\rightarrow 0$ limit. By performing a renormalization-group study of
this last theory we show that the gravitational coupling constant $G$
has
a non-trivial, ultraviolet stable fixed point (asymptotic freedom) and
that the dilatonic coupling functions (including the dilatonic
potential) exhibit also a real, non-trivial fixed point. At such point
the
theory represents a standard charged string-inspired model. Stability
and gauge dependence of the fixed-point solution is discussed. It is
shown that all these properties remain valid in a dilatonic-Yang-Mills
theory with $n$ scalars and $m$ spinors, that has the UF stable
fixed
point $G^* = 3\epsilon (48+12N-m-2n)^{-1}$. In addition, it is seen
that
by increasing $N$ (number of gauge fields) the matter central charge
$C=n + m/2$ ($0<C<24+6N$)   can be increased correspondingly (in pure
dilatonic gravity $0<C<24$).

\vspace{4mm}


\newpage

\section{Introduction}

The recent efforts that are being invested in the study of
2-dimensional quantum gravity ---in particular dilatonic
gravity, with or without matter \cite{14}-\cite{8}--- have a
variety of motivations: from the elementary fact that it is much easier
to study quantum gravity (QG) in two rather than in four dimensions,
to the much more fundamental reason, that such kind of theories
appear  naturally as string-inspired models. However,
first indications that some of the hard problems of 4D QG ---as
those associated with
the Hawking radiation and black-hole evaporation \cite{14,17}
(see \cite{19} for a review)---
might be much easier to understand in frames of 2-dimensional
dilatonic gravity, appear now not to be quite correct. The
impression nowadays is that 2-dimensional dilatonic gravity
does not look any more as a very `simple' toy model. More
effort should be invested in the study of dilatonic gravity
in exactly (or near) two dimensions, having always in mind
its subsequent generalization to higher dimensions.

As is well accepted, 2-dimensional Einstein gravity is no more of
ultimate interest, since this action represents a topological
invariant in two dimensions. However, it is quite an old idea
\cite{5} to try to study Einstein gravity in
$2+\epsilon$ dimensions, what might help to remedy this shortcoming.
The gravitational coupling constant in such a theory ---near
two dimensions--- shows an asymptotically free behavior \cite{5},
what can actually be very exciting in attempts to solve the
problem
of non-renormalizability of 4-dimensional Einstein gravity
\cite{11}.
Unfortunately, it was shown that Einstein gravity in
$2+\epsilon$  dimensions is on its turn a non-renormalizable
theory
\cite{20} (in other words, it has the oversubtraction problem
\cite{21}).

Furthermore, dynamical triangulations in more than two
dimensions (see, for example, \cite{22,28}) have clearly shown the
existence of a phase transition to a strong-coupling phase of
similar nature as 2-dimensional quantum gravity \cite{23}.
Because
of this,  of considerable interest is still to try to
construct
a consistent theory of QG  in  $2+\epsilon$  dimensions (having in
mind
a subsequent continuation of $\epsilon$ to $\epsilon =1$ or
              $\epsilon =2$).

It has been suggested recently \cite{6}, that a thing to do would be
to study dilatonic gravity near two dimensions (similarly,
it
was the idea in ref. \cite{13} to consider Einstein gravity with
a
conformal scalar field near two dimensions).
Notice, however, that unlike
Einstein gravity, dilatonic gravity possesses a smooth $\epsilon
\rightarrow 0$ limit. This difference manifests itself in the
fact
that dilatonic gravity ---which after a proper field definition
can be presented as an Einsteinian theory with a scalar field--- is
{\it not} equivalent to Einstein gravity in {\it exactly}
 two dimensions. Moreover, dilatonic gravity is renormalizable
and,
hence, it does not suffer from the oversubtraction problem. Even
more
\cite{6}, the gravitational coupling constant in this theory also has an
ultraviolet stable fixed point (asymptotic freedom) for $\epsilon
>0$
and $n<24$ ($n$ is the number of scalars or matter central
charge).
Thus, the matter central charge in dilatonic gravity is bounded,
as
it also happens in Einstein's theory.

In the present paper we consider dilatonic-Maxwell gravity with
scalar
matter in $2+\epsilon$ dimensions. This theory, which includes a
dilatonic potential and a Maxwel term with an arbitrary
dilatonic-vector
coupling function, may be considered as a toy model for
unification of
gravity with matter (scalar and vector fields). We will show that
in the
ultraviolet stable fixed point the gravitational coupling
constant has
the value $G^* = 3\epsilon /[2(30-n)]$ and, hence, owing to the
contribution of the vectors, the matter central charge of our
universe
in such a model can be naturally increased from $0<n<24$ (for
pure
dilatonic gravity) to $0<n<30$ (for dilatonic-Maxwell gravity). With
this,
we will show explicitly that it is possible to have an even wider
window for the matter central charge, provided we consider
non-abelian
gauge fields. We will also present an evaluation of the corresponding,
generalized beta functions, and their fixed-point solutions will
be
found. At the fixed point the theory can be cast under the form
of
a standard, string-inspired model with a dilatonic potential of
Liouville type.

The paper is organized as follows. In the next section we discuss
the classical action of dilatonic-Maxwell gravity with matter and
we derive the corresponding one-loop effective action. Section 3
is devoted to the study of the renormalization of the model and,
in particular, to the derivation of the beta functions (to first
order in the gravitational coupling constant $G$). The
fixed-point
solutions of the renormalization group (RG) equations together
with
a detailed analysis of their stability are presented in section
4.
The form of the action at the fixed points is discussed. In
section
5 we investigate the issue of gauge dependence of the position of the
fixed point for the dilatonic coupling function. Finally, in
section 6 we present the conclusions of our investigation and an
outlook.
\bs

\section{Dilatonic-Maxwell gravity with matter in $2+\epsilon$
dimensions}

In this section we are going to consider a theory of dilatonic
gravity
interacting with scalars and vectors via dilatonic couplings in
$2+\epsilon$ dimensions. This theory can be considered as a toy
model
for a theory of unified gravity with matter (since fermions can
be
included without any problem). The action is
 \beq
S = \int d^d x\,  \sqrt{-g} \biggl[{1 \over 2} Z(\phi) g^{\mu\nu}
\partial_\mu \phi\partial_\nu \phi + {\mu^\epsilon \over 16\pi
G} R C(\phi )+ V(\phi)
- {1 \over 2} f(\phi) g^{\mu\nu} \partial_\mu \chi_i \partial_\nu
\chi^i+ {1 \over 4} f_1(\phi) F_{\mu\nu}^2 \biggr], \label{1}
\eeq
where $g_{\mu\nu}$ is the ($2+\epsilon$)-dimensional metric, $R$
the corresponding curvature, $\chi_i$ are scalars ($i=1,2,
\ldots,
n$), $F_{\mu\nu} = \nabla_\mu A_\nu - \nabla_\nu A_\mu$, with
$A_\mu$ a vector, and where the smooth functions $Z(\phi)$,
$C(\phi)$, $V(\phi)$, $f(\phi)$ and $f_1(\phi)$ describe the
dilatonic interactions. Notice that $V(\phi)$ is a dimensional
function and it is therefore convenient to redefine $V
\rightarrow m^2V$, where $m$ is some parameter with dimensions of
mass. Notice that in the absence of an electromagnetic sector
the action (\ref{1}) can be easily presented under the form
of a non-linear sigma model \cite{24}. Then the $\epsilon
\rightarrow 0$  limit of the theory (\ref{1}) can be
discussed in a string effective action manner \cite{25}.

One can easily prove that the above theory (\ref{1}) is
renormalizable in a generalized sense. The one-loop
counterterms corresponding to (\ref{1}) in two dimensions have
been calculated in refs. \cite{1,2}. Our purpose here will be  to
study in some detail the renormalization structure of (\ref{1})
in $2+\epsilon$ dimensions and in particular the
corresponding renormalization
group equations. The very remarkable point that gives sense to
this study is the fact that the theory (\ref{1}) (unlike Einstein
gravity \cite{5}) has a smooth limit for $\epsilon \rightarrow
0$. This property allows for the possibility to study the behavior
of (\ref{1}) in $2+\epsilon$ dimensions by simply using the
counterterms  calculated already in 2 dimensions ---in close
analogy with quantum field theory in frames of the
$\epsilon$-expansion technique (for a review see \cite{3}
and \cite{4}).

Before we start working with action (\ref{1}) we will perform
some simplifications. First, motivated by the form
of the string effective action, we choose the dilatonic
couplings in (\ref{1}) as the exponents of convenient dilatonic
functions. Using the rigid rescaling of scalars and vectors:
\beq
\chi_i \longrightarrow a_1 \chi_i, \ \ \ \ \ \
A_\mu \longrightarrow a_2 A_\mu, \label{2} \eeq
with $a_1$ and $a_2$ constants, one can always normalize the
dilatonic couplings in such a manner that
\beq
f(0) =1, \ \ \ \ \ \ f_1(0)=1. \label{3} \eeq
Second, as has been done in refs. \cite{7,8}, we can also use a
local Weyl transformation of the metric of the form
\beq
g_{\mu\nu} \longrightarrow e^{-2\sigma (\phi)} \, g_{\mu\nu} ,
 \label{4} \eeq
in order to simplify the dilatonic sector of (\ref{1}). In
particular, one can use (\ref{4}) of a specific form such that
$Z(\phi)=0$. Finally, we can  perform a transformation of the
dilaton field in order to simplify the function $C(\phi)$, namely
to reduce it to the form $C(\phi)=e^{-2\phi}$. With all this taken
into account, the action (\ref{1}) can be written as follows
 \beq
S = \int d^d x\,  \sqrt{-g}\,  \biggl[ {\mu^\epsilon \over 16\pi
G}
R e^{-2\phi}
- {1 \over 2} g^{\mu\nu} \partial_\mu \chi_i \partial_\nu \chi^i
e^{-2\Phi (\phi)}+ \mu^\epsilon m^2 e^{-V(\phi )}+ {1 \over 4}
e^{-f_2(\phi)} F_{\mu\nu}^2 \biggr], \label{5}
\eeq
being $\mu$ a mass parameter and where we have choosen the
dilatonic couplings so that $\Phi (0) =f_2(0)=V(0) =0$. The
first two terms in (\ref{5}) correspond to the dilatonic gravity
action in ref. \cite{6}, which was considered in $2+\epsilon$
dimensions. With our choice of (\ref{5}) the zero modes of the
dilatonic couplings are fixed with the help of reference
operators in the gravitational sector \cite{6} and with the
renormalization (by constants) of the scalar, vector, and mass
$m^2$ in the matter sector.

We can now start the study of the divergences of the action
(\ref{5}). The study of the one-loop divergences of dilatonic
gravity in the covariant formalism, initiated in ref. \cite{9}, has been
continued
in refs. \cite{7,8,1,2,10,12}. It is by now well under control, and we
do not consider it necessary to repeat the details of such a
calculation for the case of the model (\ref{5}). Let us just
recall that it is based on the t'Hooft-Veltman prescription
\cite{11}. The gauge-fixing conditions that are most convenient
to use are the following. The covariant gauge-fixing action is
\beq
S_{gf} = -  {\mu^\epsilon \over 32\pi G} \int d^d x\,
\sqrt{-g}\,
 g_{\mu\nu} \chi^\mu \chi^\nu  e^{-2\phi}, \label{6}
\eeq
where
\beq
\chi^\mu =   \nabla_\nu \bar{h}^{\mu\nu} + 2 \nabla^\mu \varphi
\label{7}
\eeq
and $\bar{h}^{\mu\nu}$ is a traceless quantum gravitational field
and $\varphi$ a quantum scalar field, in the background field
method.
In the electromagnetic sector the gauge-fixing action is choosen
as
\beq
S_{L} = \int d^d x\,  \sqrt{-g}\,
 (\nabla_\mu Q^\mu)^2  e^{-f_2(\phi)}, \label{8}
\eeq
where $Q_\mu$ is a quantum vector field. Notice that for the
background fields we  use the same notations as for the classical
fields in (\ref{5}).

The calculation of the one-loop effective action corresponding to
(\ref{5}) in the gauges (\ref{6}) and (\ref{8}) can be performed
in close analogy with the one in refs. \cite{1,2} (which was
actually carried out for a more general theory in two
dimensions). Owing to the smooth behavior of (\ref{5}) for
$\epsilon \rightarrow 0$, the divergences of (\ref{5}) can be
also calculated in exactly two dimensions, what will also provide
the result for $2+\epsilon$ dimensions. After some algebra, we
get
\bea
\Gamma_{div} &=& \frac{1}{4\pi \epsilon}  \int d^d x\,
\sqrt{-g}\,
\left\{ \frac{30-n}{6} R +m^2 e^{2\phi - V(\phi)} 16\pi G [2 +
V'(\phi)] -  g^{\mu\nu} \partial_\mu \phi \partial_\nu \phi [8-
n\Phi'(\phi)^2] \right. \nn \\
&& + \left.  {4\pi G \over \mu^\epsilon} e^{2\phi-f_2(\phi)}
F_{\mu\nu}^2 [f_2'(\phi) -2 ] \right\},
\label{9}
\eea
where $\epsilon =d-2$.
A few remarks about the comparison of (\ref{9}) in some
particular cases with results that have appeared in the
literature are in order. First, in the absence of scalars and
vectors (pure dilatonic gravity), we obtain
\beq
\Gamma_{div} = \frac{1}{4\pi \epsilon}  \int d^d x\,  \sqrt{-g}\,
\left\{ 4R +m^2 e^{2\phi - V(\phi)} 16\pi G [2 + V'(\phi)] - 8
g^{\mu\nu} \partial_\mu \phi \partial_\nu \phi \right\}.
\label{10}
\eeq
This expression coincides with the results of refs.
\cite{9,7,8,1,2,10,12} in the same covariant gauge. Moreover, in
ref. \cite{8}, $\Gamma_{div}$ eq. (\ref{10}) has been calculated
in a one-parameter dependent gauge, what gives an additional
check of (\ref{10}) at the value of the gauge parameter
corresponding to the gauge choice (\ref{6}). For dilatonic
gravity with a Maxwell sector, (\ref{9}) coincides with the
results in \cite{1,2,10}. Notice that the calculation of
$\Gamma_{div}$ eq. (\ref{9}) for the theory (\ref{5}) without
vectors and for $e^{-V(\phi)}\equiv 0$ has been done in ref.
\cite{6}, in the ($2+\epsilon$)-dimensional formalism but in a
slightly different gauge. When comparing the results, there is
here a difference with the coefficient of the $g^{\mu\nu}
\partial_\mu \phi \partial_\nu \phi$ term in (\ref{9}). On one
hand, we get a different contribution from the scalar dilaton
coupling, what may be due to the slightly different gauges. On
the other, instead of our coefficient 8 there appears in \cite{6}
a coefficient 4, what might be an indication of some mistake,
or maybe due to the use of the inconvenient background field
method
---since these authors also employ a gauge of harmonic type and do not
agree with the results of \cite{1,7,8,10} either. Later we will
discuss the gauge dependence of  $\Gamma_{div}$.
\bs

\section{Beta functions and renormalization group analysis}

We turn now to the study of the RG corresponding to our model in
$2+\epsilon$ dimensions. We will follow here the approach of ref.
\cite{6}, which is actually based on earlier considerations in
\cite{5}. The main idea of the whole approach is to use $G$ as
the coupling constant in perturbation theory and work at some
fixed power of $G$.
The counterterm which follows from (\ref{9}) can be written as
\beq
\Gamma_{count} =- \mu^\epsilon  \int d^d x\,  \sqrt{-g}\,
\left[
RA_1(\phi) + g^{\mu\nu} \partial_\mu \phi \partial_\nu \phi
A_2(\phi) + m^2A_3(\phi) +  \frac{\mu^{-\epsilon}}{4}
F_{\mu\nu}^2 A_4(\phi) \right],
\label{11}
\eeq
where
\bea
A_1(\phi) &=& \frac{30-n}{24\pi\epsilon} \equiv A_1, \nn \\
A_2(\phi) &=& \frac{1}{4\pi\epsilon} [n\Phi'(\phi)^2-8] , \nn
\\
A_3(\phi) &=& \frac{4G}{\epsilon}[2 + V'(\phi)]e^{2\phi -V(\phi)}
\equiv e^{-V(\phi)} \wt{A}_3 (\phi), \nn \\
A_4(\phi) &=& \frac{4G}{\epsilon}[f_2'(\phi)-2]e^{2\phi -
f_2(\phi)} \equiv e^{-f_2(\phi)} \wt{A}_4 (\phi).
\label{12}
\eea
The structure of the counterterms shows that the gravitational
coupling is a really nice choice for parameter of the
perturbation theory.

By looking carefully at the relations (\ref{12}) we observe that
our theory is one-loop finite provided the following conditions
are fulfilled:
\beq
n=30, \ \ \ \ \Phi (\phi) = \sqrt{\frac{8}{n}} \, \phi, \ \ \ \
V(\phi) =-2 \phi, \ \ \ \ f_2(\phi) =2\phi.
\label{12a}
\eeq
(Notice, however, that perturbative finiteness is a more natural
property in 2-dimensional supergravity \cite{26}).

Renormalization is now performed in the standard way
 \bea
S& =& S_{cl} + S_{count} =\int d^d x\,  \sqrt{-g}\,  \biggl[ {1
\over 16\pi G_0} R e^{-2\phi_0}
- {1 \over 2} g_0^{\mu\nu} \partial_\mu \chi_{0i} \partial_\nu
\chi_0^i e^{-2\Phi_0 (\phi_0)} \nn \\
&&+  \mu^\epsilon m_0^2 e^{-V_0(\phi_0 )}+ {1 \over 4} e^{-
f_{02}(\phi_0)} g_0^{\mu\alpha}g_0^{\nu\beta}F_{0\mu\nu}
F_{0\alpha\beta} \biggr]. \label{13}
\eea
The renormalization transformations can be defined as follows
\bea
&& \phi_0 = \phi + f(\phi), \ \ \ g_{0\mu\nu} =g_{\mu\nu}e^{-
2\Lambda (\phi)}, \ \ \ \Phi_0 (\phi_0)= \Phi(\phi) +
F(\phi), \ \ \ V_0 (\phi_0)= V(\phi) + F_V(\phi), \nn \\
&&  f_{02} (\phi_0)= f_{2}(\phi) + F_f(\phi), \ \ \ \chi_{0i} =
Z_\chi^{1/2} \chi, \ \ \ A_{0\mu} = Z_A^{1/2} A_\mu, \ \ \ m_0^2
=Z_m m^2. \label{14}
\eea
The functions $\Lambda$, $f$, $F$,... are chosen so that
$\Lambda(0)$, $f(0)$, $F(0)$,... can be set equal to zero.
Substituting the renormalization transformations (\ref{14}) into
the renormalized action (\ref{13}) (in close analogy with ref.
\cite{6}), we obtain
 \bea
S_0& =& \int d^d x\,  \sqrt{-g}\,  \biggl\{ {1 \over 16\pi G_0} R
e^{-2\phi-2f(\phi)-\epsilon \Lambda (\phi)} \nn \\ &&
+ {\epsilon +1 \over 16\pi G_0} [4\Lambda'(\phi) + \epsilon
\Lambda'(\phi)^2+
4f'(\phi)\Lambda'(\phi)]e^{-2\phi-2f(\phi)-
\epsilon \Lambda (\phi)} g^{\mu\nu} \partial_\mu \phi
\partial_\nu \phi \nn \\
&&- {1 \over 2}Z_\chi g^{\mu\nu} \partial_\mu \chi_{i}
\partial_\nu
\chi^i e^{-2\Phi(\phi)-2F(\phi)-\epsilon \Lambda (\phi)}+  \mu^\epsilon
m^2 Z_m e^{-V(\phi )-F_V(\phi)-(2+\epsilon) \Lambda (\phi)} \nn
\\ &&+ {Z_A \over 4} e^{-f_{2}(\phi)-F_f (\phi)+(2-\epsilon)
\Lambda (\phi) } F_{\mu\nu}^2 \biggr\}. \label{15}
\eea
 From here, one can easily obtain
\bea
&& {1 \over 16\pi G_0}
e^{-2\phi-2f(\phi)-\epsilon \Lambda (\phi)} =   \mu^\epsilon
\left( {1 \over 16\pi G}  e^{-2\phi} -A_1 \right), \nn \\
&& {\epsilon +1 \over 16\pi G_0} [4\Lambda'(\phi) + \epsilon
\Lambda'(\phi)^2+ 4\epsilon f'(\phi)\Lambda'(\phi)]
e^{-2\phi-2f(\phi)-\epsilon \Lambda (\phi)} =  - \mu^\epsilon
A_2(\phi), \nn \\
&& Z_\chi e^{-2\Phi(\phi)-2F (\phi)-\epsilon
\Lambda (\phi) }= e^{-2\Phi(\phi)}, \nn \\
&& Z_A e^{-f_{2}(\phi)-F_f (\phi)+(2-\epsilon)
\Lambda (\phi) }= e^{-f_{2}(\phi)} [1- \wt{A}_4 (\phi)], \nn \\
&& Z_m e^{-V(\phi)-F_V (\phi)-(2+\epsilon)
\Lambda (\phi) }= e^{-V(\phi)} [1- \wt{A}_3 (\phi)]. \label{16}
\eea
Working to leading order in $G$ (notice that the functions $f$,
$\Lambda$, $F$,..., $F_f$ are of first order in $G$), we may
follow ref. \cite{6} and obtain the renormalized parameters as
\bea
&& G_0^{-1}= \mu^\epsilon (G^{-1} -16 \pi A_1), \ \ \ Z_\chi =1,
\
\ \ \Lambda (\phi) = - \frac{4\pi G}{\epsilon +1} \int_0^\phi
d\phi' \, e^{2\phi'}A_2(\phi'), \ \ \ \epsilon \Lambda (\phi) =-
2F(\phi), \nn \\
&& f(\phi) = 8\pi A_1 G(e^{2\phi} -1)+\frac{2\pi\epsilon
G}{\epsilon +1} \int_0^\phi d\phi' \, e^{2\phi'}A_2(\phi'), \ \ \
Z_A =1-A_4 (0), \ \ \ Z_m =1-A_3 (0), \nn \\
&& F_f (\phi)= (2-\epsilon) \Lambda (\phi) + \wt{A}_4(\phi) -
\wt{A}_4(0), \ \ \ F_V (\phi)=-(2+\epsilon) \Lambda (\phi) +
\wt{A}_3(\phi) -\wt{A}_3(0). \label{17}
\eea
Hence, we see that the zero modes of the functions $f$, $f_2$ and
$V$ are indeed controlled by a constant renormalization of the
vector, the scalar and the mass.

We can now turn to the evaluation of the beta functions. We will
list below only the ones corresponding to the dilatonic couplings
and gravitational constant (for the purely dilatonic case they
were given in ref. \cite{6})
\bea
\beta_G &=& \mu {\partial G \over \partial \mu} = \epsilon G - 16
\pi \epsilon A_1 G^2, \nn \\
&& \mu {\partial \Lambda (\phi_0) \over \partial \mu} =  -
\frac{4\pi
\epsilon G}{\epsilon +1} \int_0^\phi d\phi' \,
e^{2\phi'}A_2(\phi'), \nn \\
\beta_\Phi (\phi_0) &=& \mu {\partial \Phi (\phi_0)  \over
\partial \mu} = 8\pi \epsilon A_1 G (e^{2\phi_0}-1)\Phi'
(\phi_0) \nn \\
&&+ \frac{2\pi \epsilon^2 G}{\epsilon +1} [\Phi' (\phi_0) -1]
\int_0^{\phi_0} d\phi' \, e^{2\phi'}A_2(\phi'), \nn \\
\beta_{f_2} (\phi_0) &=& \mu {\partial f_2 (\phi_0)  \over
\partial \mu} = 8\pi \epsilon A_1 G (e^{2\phi_0}-1) f_2' (\phi_0)
\nn \\
&&+ \frac{2\pi \epsilon G}{\epsilon +1} [\epsilon f_2' (\phi_0)
+2(2-\epsilon)] \int_0^{\phi_0} d\phi' \, e^{2\phi'}A_2(\phi')-
\epsilon [\wt{A}_4(\phi_0) -\wt{A}_4(0)] , \nn \\
\beta_V (\phi_0) &=& \mu {\partial V (\phi_0)  \over \partial
\mu}
= 8\pi \epsilon A_1 G (e^{2\phi_0}-1) V' (\phi_0) \nn \\
&&+ \frac{2\pi \epsilon G}{\epsilon +1} [\epsilon V' (\phi_0)
-2(2+\epsilon)] \int_0^{\phi_0} d\phi' \, e^{2\phi'}A_2(\phi')-
\epsilon [\wt{A}_3(\phi_0) -\wt{A}_3(0)]. \label{18}
\eea
Similarly, one can obtain the $\gamma$-functions for the fields
and mass parameter, which are however of less importance to us,
due to the fact that they correspond to non-essential couplings.
Notice also that  the only conformal mode of the gravitational
field is renormalized in $2+\epsilon$ dimensions, what is quite
well known \cite{5,6} (for a study of conformal factor dynamics in four
dimensions, see \cite{29}). \bs

\section{Fixed points}

The $\beta$-function for the gravitational constant has the form
(\ref{18})
\beq
\beta_G = \epsilon G - \frac{2(30-n)}{3} G^2. \label{19}
\eeq
Hence, for $n<30$ we obtain the infrared stable fixed point
$G=0$. There is also an ultraviolet stable fixed point as in \cite{5}
\beq
G^* = \frac{3\epsilon}{2(30-n)}. \label{20}
\eeq
The theory is asymptotically free in the ultraviolet limit. The
inclusion of vector fields has increased the matter central
charge of our universe, when compared with the cases of pure
dilatonic gravity \cite{6}, Einstein gravity \cite{5}, or
Einstein gravity with a conformal scalar \cite{13}. Our theory
admits more matter than any of these previous models.

We start the search for fixed-point solutions corresponding to
the dilatonic couplings. To this end we choose the following {\it
Ansatz}:
\beq
\Phi (\phi) = \lambda \phi, \ \ \ \
f_2 (\phi) = \lambda_f \phi, \ \ \ \
V (\phi) = \lambda_V \phi.
\label{21}
\eeq
In that case the $\beta$-functions are (we study the functional
dependence of $\beta$-functions of the type (\ref{21}))
\bea
\beta_\Phi &=& G(e^{2\phi} -1) \left[ \frac{30-n}{3} \lambda +
\frac{\epsilon}{4(1+\epsilon)} (\lambda -1) (n\lambda^2-8)
\right], \nn \\
\beta_{f_2} &=& G(e^{2\phi} -1) \left[ \frac{30-n}{3} \lambda_f +
\frac{n\lambda^2-8}{4(1+\epsilon)} (\epsilon\lambda_f+4 -
2\epsilon) -4\lambda_f +8 \right], \nn \\
\beta_V &=& G(e^{2\phi} -1) \left[ \frac{30-n}{3} \lambda_V +
\frac{n\lambda^2-8}{4(1+\epsilon)} (\epsilon\lambda_V-4 -
2\epsilon) -4\lambda_V -8 \right].
\label{22}
\eea

Equating these $\beta$-functions to zero we easily get the fixed-point
solutions.
For the dilatonic coupling, we find
\beq
\lambda^* = -\frac{6\epsilon}{30-n} + {\cal O} (\epsilon^2).
\label{23}
\eeq
There are also imaginary oscillating solutions for $\lambda \sim
\epsilon^{-1/2}$, which have been mentioned in ref. \cite{6} and
which are not physical solutions.

Using (\ref{23}), for the Maxwell-dilatonic coupling and dilatonic
potential, we obtain
\beq
\lambda_f^* = -\frac{36\epsilon}{18-n} + {\cal O} (\epsilon^2), \
\
\ \ \ \ \lambda_V^* = \frac{12\epsilon}{18-n} + {\cal
O} (\epsilon^2) \label{24}
\eeq
The fixed-point solutions (\ref{24}) are of the same nature as
solution (\ref{23}). Notice, however, that the denominator in
(\ref{24}) is different from the denominator for the case of a
purely dilatonic sector (\ref{23}). For the existence of the solutions
(\ref{24}) a new limitation appears: $n\neq 18$. Under a discontinuous
transition through the point $n=18$, the sign of the fixed points in
(\ref{24}) changes.

It is interesting to see in which way the model (\ref{5}) can be
rewritten at the fixed point. In particular, we will perform a
Weyl
rescaling (it is non-singular) in the manner suggested in ref.
\cite{6}
\beq
g_{\mu\nu} \longrightarrow g_{\mu\nu}  \, \exp \left( \frac{4
\lambda^*}{\epsilon} \phi \right). \label{25}
\eeq
Then, the classical action becomes
 \bea
S& =& \int d^d x\,  \sqrt{-g}\,  \biggl\{ {\mu^\epsilon \over
16\pi G^*}
e^{-2(1-\lambda^*)\phi} \left[ R - \frac{4(1+\epsilon)}{\epsilon}
\lambda^* (2- \lambda^*) \right] g^{\mu\nu} \partial_\mu \phi
\partial_\nu \phi  \nn \\
&&- {1 \over 2} g^{\mu\nu} \partial_\mu \chi_{i} \partial_\nu
\chi^i +  \mu^\epsilon
m^2  e^{(2\lambda^* +4\lambda^*/\epsilon- \lambda^*_V)\phi}
+\frac{1}{4} F_{\mu\nu}^2 e^{(2\lambda^* -4\lambda^*/\epsilon-
\lambda^*_f)\phi} \biggr\}. \label{26}
\eea
As one can see from expression (\ref{26}), without the
electromagnetic sector and setting  $m^2=0$, it describes the
CGHS action \cite{14} (at the limit $\epsilon \rightarrow 0$ is finite
and scalars become non-interacting with the dilaton). As it stands,
action (\ref{26}) is of a  similar form as dilaton-Maxwell
gravity \cite{15,2} with the Liouville potential. Notice that
such  kind of dilaton-Maxwell gravity (which can also be
considered as a charged string-inspired model \cite{15}) admits charged
black hole solutions with multiple horizons, being in
this sense analogous to four or higher-dimensional
Einstein-Maxwell theories \cite{16}. Different forms of dilatonic
gravity can be easily obtained too, by transforming the metric and the
dilaton (see the appendix). For example, by transforming $g_{\mu\nu}
\rightarrow \exp
(\frac{2\epsilon}{2- \epsilon} \lambda_f^* \phi)  \, g_{\mu\nu}$, we can
present the theory at the critical point as having a free (i.e.,
non-interacting with the dilaton) Maxwell sector.

Notice also that, in order to fix the scale of the metric ---what
is certainly necessary for discussing the renormalization of the
gravitational constants \cite{5}--- one can use the same reference
operator as in ref. \cite{6}, e.g. a combination of the trace of
the metric and of the dilaton. As has been already explained in
detail in \cite{6}, this choice is enough to fix the scale of
$g_{\mu\nu}$ and the origin of $\phi$.

One can now study the
stability of the fixed points (\ref{20}), (\ref{23}), (\ref{24}),
along the same lines as in ref. \cite{6}. Now we can perform
variations along four different trajectories, but concerning the
existence of fixed points the situation is pretty similar to that
in ref. \cite{6}. In fact, a careful analysis of the the
beta-functions (\ref{18}) for the linear Ansatz (\ref{21}) shows
that the two last equations (i.e. those for $f_2$ and $V$) do not
produce a new multiplicity of solutions. In other words, for each
value of $G^*$ and $\lambda^*$ we just have one single value of
$\lambda_f^*$ and one of $\lambda^*_V$, that complete the four
coordinates of the fixed point. For $\lambda^*$ we obtain three
distinct solutions: the real one (\ref{23}) and two purely
imaginary ones, of order $\epsilon^{-1/2}$, namely
\beq
\lambda^*_\pm = \pm 2i\sqrt{\frac{30-n}{3n\epsilon}} + {\cal O}
(\epsilon^0), \label{261}
\eeq
that correspond to highly oscillating dilaton couplings and seem
not to have any sensible meaning.

Expanding the beta functions near the (only real) fixed point, in
the way
\beq
G=G^*+ \delta G, \ \ \ \ \Phi=\lambda^*\phi+ \delta \Phi, \ \ \ \
f_2=\lambda_f^* \phi+ \delta f_2, \ \ \ \ V=\lambda^*_V \phi +
\delta V, \label{262}
\eeq
and assuming all the fluctuations to be small, we obtain
\bea
\delta \beta_G &=&- \epsilon \delta G, \nn \\
\delta \beta_\Phi &=&\frac{ \epsilon}{2} \left(e^{2\phi} -
1\right) \frac{d}{d\phi} \delta \Phi + {\cal O} (\epsilon^2), \nn
\\
\delta \beta_{f_2}&=&\frac{ \epsilon}{2} \left( \frac{18-n}{30-n}
e^{2\phi} -1\right) \frac{d}{d\phi} \delta f_2 + {\cal O}
(\epsilon^2), \nn \\
\delta \beta_V&=&\frac{ \epsilon}{2} \left( \frac{18-n}{30-n}
e^{2\phi} -1\right) \frac{d}{d\phi} \delta V + {\cal O}
(\epsilon^2). \label{263}
\eea
As observed in ref. \cite{6}, we may take $e^\phi$ to play the
role of loop expansion parameter, and restrict ourselves to the
region $e^{2\phi} \leq 1$, that is $-\infty < \phi \leq 0$. The
change of variables
\beq
\eta_1 =\ln (e^{-2\phi} -1), \ \ \ \ \ \
\eta_2 =\ln \left[ \frac{30-n}{12} \, \left( e^{-2\phi} -
\frac{18-n}{30-n}\right) \right], \label{264}
\eeq
transform this region into the following ones
\bea
\phi =0, & \eta_1 \rightarrow -\infty,  & \eta_2 =0,  \nn \\
\phi \rightarrow -\infty, & \eta_1 \rightarrow +\infty,  & \eta_2
\rightarrow +\infty,  \label{265}
\eea
respectively. They simplify expressions (\ref{263}), which now
read:
\bea
\delta \beta_G &=&- \epsilon \delta G, \nn \\
\delta \beta_\Phi &=& \epsilon  \frac{d}{d\eta_1} \delta \Phi +
{\cal O} (\epsilon^2), \nn
\\
\delta \beta_{f_2}&=& \epsilon  \frac{d}{d\eta_2} \delta f_2 +
{\cal O} (\epsilon^2), \nn
\\
\delta \beta_V &=& \epsilon  \frac{d}{d\eta_2} \delta V + {\cal
O} (\epsilon^2).
 \label{266}
\eea
The eigenfunctions corresponding to the new differential
operators on the right hand side are, respectively,
\bea
\delta \Phi &\sim& e^{\alpha_1 \eta_1} = (e^{-2\phi} -
1)^{\alpha_1}, \nn \\
\delta f_2 &\sim & e^{\alpha_2 \eta_2} = \left[ \frac{30-n}{12}
\, \left( e^{-2\phi} -\frac{18-n}{30-n}\right) \right]^{\alpha_2},
\nn \\
\delta V &\sim & e^{\alpha_3 \eta_2} = \left[ \frac{30-n}{12}
\, \left( e^{-2\phi} -\frac{18-n}{30-n}\right) \right]^{\alpha_3},
\label{267}
\eea
and the corresponding eigenvalues are
\beq
\delta \beta_\Phi =\epsilon \alpha_1 \delta \Phi, \ \ \ \
\delta \beta_{f_2}=\epsilon \alpha_2 \delta f_2, \ \ \ \
\delta \beta_V=\epsilon \alpha_3 \delta V.
 \label{268}
\eeq
By imposing the initial condition $\delta \Phi (\phi =0) =0$ we
see that $\alpha_1 >0$.
Hence, we observe that the fixed point for $G$ is not ultraviolet stable
in the direction $\delta \Phi$. At the same time, within such a picture
it is clear
that we can always choose $\alpha_2$ and $\alpha_3$ to be negative
---since, for the other two directions, $f_2$ and
$V$, there is no way to make $\delta f_2 =0$ nor $\delta V =0$ at
$\phi =0$. (By means of an adequate choice
of arbitrary constant, we can choose $\delta f_2 (\phi =0) $ and
$\delta V (\phi =0) $ as small as we like, but never zero ---with
the choice above we have set these two values equal to 1).
Therefore, the fixed points for $f_2$ and $V$ are ultraviolet stable in
the direction $\delta \Phi$ but are always infrared unstable in this
direction (since they are never zero at $\phi =0$). In this sense the
theory (four RG functions) possesses a saddle fixed point.
 \bs

\section{Gauge dependence and fixed points}

It is of interest to discuss in some detail how the results of
the previous sections would change if we made a different choice
of the gauge condition. This question is not trivial at all, as
was already mentioned in \cite{8,9}, where it was shown that at
some gauges dilatonic gravity could be rendered one-loop finite
(except for the conformal anomaly term). However, explicit
calculation of the divergences for dilatonic gravity with matter
in a parameter-dependent gauge are extremely cumbersome. Hence,
we shall here consider the simplified model of ref. \cite{8}
\beq
S= \int d^d x\,  \sqrt{-g}\,  \left\{ {1 \over
2}  g^{\mu\nu} \partial_\mu \varphi
\partial_\nu \varphi +C_1 \varphi R+ f_3(\varphi) g^{\mu\nu}
\partial_\mu \chi_{i} \partial_\nu \chi^i\right\},
\label{27}
\eeq
where $f_3(\varphi) $ is some dilatonic coupling and $C_1$ some
constant. The one-loop divergences of the theory (\ref{27}) can
be studied in the one-parameter dependent gauge
\bea
S_{gf} &=& - \frac{C_1}{2\alpha} \int d^d x\,  \sqrt{-g}\,
\chi_\mu \varphi \chi^\mu, \nn \\
\chi_\mu &=& \nabla_\mu h^\nu_{\ \mu} - \frac{1}{2} \nabla_\mu h
- \frac{\alpha}{\varphi} \partial_\mu \wt{\varphi} + X(\varphi)
\wt{\varphi}  \partial_\mu \varphi \nn \\
&&+ h_{\rho\sigma} \left[ Y_1 (\varphi) \left( \delta^\rho_\mu
\nabla^\sigma +\delta^\sigma_\mu \nabla^\rho \right) \varphi +
Y_2 (\varphi)  g^{\rho\sigma} \partial_\mu \varphi \right],
\label{28}
\eea
where $\varphi$  is a background field ($\varphi \rightarrow
\varphi +\wt{\varphi}$, $g_{\mu\nu}  \rightarrow g_{\mu\nu}
+h_{\mu\nu}$), $\wt{\varphi}$ and $h_{\mu\nu}$ are quantum
fields, $\alpha$ a gauge parameter, and $X,Y_1,Y_2$ are arbitrary
dilatonic functions.

The calculation of the one-loop effective action for the theory
(\ref{27}) in the gauge (\ref{28}) (let aside from the
$f_3(\varphi)$-dependence) has been performed in ref. \cite{8}.
The result is now
\beq
\Gamma_{div} =  \frac{1}{2\pi\epsilon} \int d^d x\,  \sqrt{-g}\,
\left\{ \frac{24-n}{12}R + \left[ - \frac{\alpha}{\varphi^2} +
nV_1 (\alpha, f_3(\varphi)) \right] g^{\mu\nu} \partial_\mu \varphi
\partial_\nu \varphi \right\}.
\label{29}
\eeq
Notice that in ref. \cite{8} only the dilatonic contribution ---
the (-$\alpha/\varphi^2$)-term--- had been found. The explicit
calculation of the function $V_1$, which depends on $f_3$ and
$\alpha$, requires a huge algebra (see \cite{8}). Fortunately,
knowledge of the precise form of $V_1$ is not necessary for our
considerations here, as we will see below.

Now, performing the background transformation in (\ref{27}) (in
exactly two dimensions)
\beq
g_{\mu\nu} \longrightarrow \exp \left( - \frac{\varphi}{4C_1}
\right) \, g_{\mu\nu}, \ \ \ \ \ \ \varphi =e^{-2\phi},
\label{30}
\eeq
we can rewrite (\ref{27}) as follows
\beq
S= \int d^d x\,  \sqrt{-g}\,  \left\{ {1 \over
16\pi G} R e^{-2\phi}+ f_3(\phi) g^{\mu\nu} \partial_\mu \chi_{i}
\partial_\nu \chi^i \right\},
\label{31}
\eeq
where $C_1$ is to be identified with $(16\pi G)^{-1}$ and
$f_3(\phi)$ with $\exp (-2 \Phi (\phi)$.
After proper transformation of the gauge-fixing Lagrangian
(\ref{28}), the $\Gamma_{div}$ corresponding to the theory
(\ref{27}) ---which is a particular case of the model
(\ref{5})--- in two dimensions is
\beq
\Gamma_{div}= \frac{1}{2\pi \epsilon} \int d^d x\,  \sqrt{-g}\,
\left\{ {24-n \over
12} R + 4 \left[ - \alpha + n V_1 (\alpha, \phi ) \right] g^{\mu\nu}
\partial_\mu \phi \partial_\nu \phi \right\}.
\label{32}
\eeq
For $\alpha =1$ the dilatonic gravity contribution in (\ref{32})
coincides with the result given in \cite{10,9,7,8} or in Eq. (\ref{10}).

Repeating the considerations of sects. 3 and 4 with $\Gamma_{div}$
(\ref{32}), we see that $A_2(\phi)$ changes  and hence
$\beta_\Phi$ changes also accordingly (but not $\beta_G$). Adding to the
central charge in (\ref{32}) the contribution from the Maxwell sector,
and looking for the fixed-point solution  (\ref{20}) and (\ref{21}), we
find
\beq
G^* = \frac{3\epsilon}{2 (30-n)}, \ \ \ \ \ \ \lambda^* =-
\frac{6\epsilon \alpha}{30-n}.
\label{33}
\eeq
The explicit form of $V_1$ (which is a combination of $\Phi$ and its
derivatives) is not necessary to find this fixed point, as it was also
the case with eq. (\ref{23})). The value of the gravitational coupling
constant at the ultraviolet fixed-point is gauge independent, as it
should be.

We realize that for any non-zero value of the gauge parameter a fixed
point $\lambda^*$ exists. (The case $\alpha =0$, which corresponds to a
gauge of Landau type, is in some sense degenerate and, hence, all the
considerations in this case must be given independently, including the
calculation of the explicit form of the function $V_1$). The change of
gauge parameter will affect the position of the fixed point through the
slope of the function $\Phi (\phi)$. However, the more physical issue of
the stability (or instability) of this fixed point will not be affected.
\bs

\section{Conclusions}

In this paper we have studied dilatonic-Maxwell gravity with matter near
two dimensions. The nice properties of this theory are: (i) its
renormalizability in $2+\epsilon$ dimensions and the asymptotically free
behavior in the ultraviolet regime shown by the gravitational coupling
constant. (ii) The fact that there is a  non-trivial fixed-point
solution for the dilatonic couplings and that, at the fixed point, the
theory may be represented in the standard form of the string-inspired
models that have been discussed recently. (iii) The increase of the
upper limit for the matter central charge ---due to the contribution of
the vector field--- from 24 (pure dilatonic gravity) or 25 (Einstein
gravity) to 30. This gives the possibility to extend the matter content
of the theory.

The investigation of the theory (\ref{5}) shows the way of considering
even more realistic toy models in $2+\epsilon$ dimensions. Indeed,
instead of the Maxwell term in (\ref{5}) we can insert its
non-abelian, Yang-Mills generalization corresponding to a gauge
potential $A_\mu^a$, $a=1,2, \ldots, N$. For simplicity, let us consider
the gauge group to be simple and compact and the structure constants
antisymmetric. Let us also add to  action (\ref{5}) the kinetic term for
$m$ fermions with a dilatonic coupling constant similar to $\Phi
(\phi)$. Such theory represents the unification of dilatonic QG with
matter (scalars, spinors and vectors) in $2+\epsilon$ dimensions.

The beta-function for the gravitational coupling constant is found to be
\beq
\beta_G = \epsilon G - \frac{1}{3} (48+ 12N - m - 2n) G^2.
\eeq
 From this expression we see that the matter central charge, $C=n+m/2$
is limited as follows:
\beq
0<C<24+6N,
\eeq
and for such a range of the central charge there is an ultraviolet
stable fixed point for the gravitational coupling constant, namely
\beq
G^* = \frac{3\epsilon}{48+12N-m-2n}, \ \ \ \ \epsilon >0.
\eeq
Hence, we still preserve asymptotic freedom in the gravitational
coupling constant. Moreover, we have now much less rigid restrictions to
the central charge: by increasing the dimension $N$ of the gauge
group we may increase the number of scalars and spinors in the theory.
The calculation of the one-loop effective action for this theory can be
carried out following ref. \cite{1} (and the second ref. of \cite{10})
and, at least qualitatively, the conclusions
about the existence of  non-trivial fixed point solutions of the form
(\ref{21}) remain true. In particular, we obtain for the above model:
\beq
\lambda^* = -\frac{12\epsilon}{48+12N-m-2n}, \ \ \
\lambda^*_f = -\frac{72\epsilon}{24+12N-m-2n}, \ \ \
\lambda^*_V = \frac{24\epsilon}{24+12N-m-2n}.
\eeq
The stability of the fixed-point solutions (as well as for the
corresponding fermion-dilaton coupling) can be studied similarly as in
sect. 4.

It is also of interest to consider supergravity models in $2+\epsilon$
dimensions. Some attempt in this direction has already been started in
ref. \cite{27}. Finally, an important issue is to study
quantum cosmology in frames of our ($2+\epsilon$)-dimensional model and,
in particular, to check carefully the claim of ref. \cite{13} that
RG considerations in $2+\epsilon$ dimensions may indeed help to solve
the spacetime singularity problem. We plan to return to some of these
questions in the near future.
 \vspace{5mm}


\noindent{\large \bf Acknowledgments}

We  would like to thank N. Sakai and Y. Tanii for very helpful
discussions.
This work has been supported by DGICYT (Spain), project Nos.
PB93-0035
and SAB93-0024, and by CIRIT (Generalitat de Catalunya).

\begin{appendix}

\renewcommand{\theequation}{{\mbox A}.\arabic{equation}}

\section{Appendix: Jackiw-Teitelboim model with matter in $2+\epsilon$
dimensions}

\setcounter{equation}{0}

One of the most popular models of 2D QG is the so-called
Jackiw-Teitelboim (JT) model \cite{18}
\beq
S_{cl}= \int d^d x\,  \sqrt{-g}\,  \left[ \mu^\epsilon e^{-2\phi} \left(
\frac{R}{16\pi G} + m^2 \right) \right].
\label{a1}
\eeq
The classical solution of this theory is
\beq
R= -16\pi G m^2
\label{a2}
\eeq
and shows the quite remarkable fact that there is only possibility for
existence of constant curvature geometries in the model. It also gives
the way to construct non-critical string theories without limitations in
the matter central charge of the theory \cite{18}.

By interaction of the theory (\ref{a1}) with scalars and vectors under
the form of a dilatonic coupling, as in (5), the property (\ref{a2}) is
lost. Dilatonic field equations become more complicated, involving also
contributions from the matter sector. By performing renormalization as
it was done before, we observe that the choice of $V(\phi)$ as in
(\ref{a1}) breakes down at the quantum level (25), since $\lambda_V^*
\neq 0$. Hence, the JT model with matter is not a non-trivial fixed
point of the RG in $2+\epsilon$ dimensions. Of course, starting from
action (5) and carring out a non-singular gauge transformation of the
metric as in (26) one can obtain in the action ---at the fixed point---
a term of the form (\ref{a1}). However, the price for this will be the
appearence of a dilatonic kinetic term, what destroys the crucial
property of the JT model (\ref{a2}) already in the sector of pure
dilatonic gravity.

\end{appendix}

\end{document}